\documentclass[astropac,article,submit,moreauthors,pdftex]{mdpi}

\firstpage{1} 
\pubvolume{xx}
\issuenum{1}
\articlenumber{5}
\pubyear{2020}
\copyrightyear{2020}
\history{Received: date; Accepted: date; Published: date}

\usepackage{comment}
\usepackage{amsmath}
\usepackage{amsthm}
\usepackage{url}
\usepackage{color,colortbl}
\definecolor{light}{rgb}{0.95,0.95,0.95}
\usepackage{booktabs}
\usepackage{tipa}
\usepackage{natbib}
\usepackage[utf8]{inputenc}

\Title{Climate Change and Astronomy: A Look at Long-Term Trends on Maunakea}

\Author{Maaike A. M. van Kooten *$^{1\dagger}$\orcidA{} and Jonathan G. Izett $^{2\dagger}$\orcidB{}} 

\AuthorNames{Maaike A. M. van Kooten and Jonathan G. Izett}

\address{%
$^{1}$ \quad Department of Astronomy and Astrophysics, University of California, Santa Cruz\\
$^{2}$ \quad Department of Ocean Sciences, University of California, Santa Cruz}

\corres{\hangafter=1 \hangindent=1.05em \hspace{-0.82em}Correspondence: mvankoot@ucsc.edu}

\abstract{Maunakea is one of the world's primary sites for astronomical observing, with multiple telescopes operating over sub-millimeter to optical wavelengths. With its summit higher than 4200 meters above sea level, Maunakea is an ideal location for astronomy with an historically dry, stable climate and minimal turbulence above the summit. Under a changing climate, however, we ask how the (above-) summit conditions may have evolved in recent decades since the site was first selected as an observatory location, and how future-proof the site might be to continued change. We use data from a range of sources, including in-situ meteorological observations, radiosonde profiles, and numerical reanalyses to construct a climatology at Maunakea over the previous 40 years. We are interested in both the meteorological conditions (e.g., wind speed and humidity), and the image quality (e.g., seeing). We find that meteorological conditions were, in general, relatively stable over the period with few statistically significant trends and with quasi-cyclical inter-annual variability in astronomically significant parameters such as temperature and precipitable water vapour. We do, however, find that maximum wind speeds have increased over the past decades, with the frequency of wind speeds above 15~m~s$^{-1}$ increasing in frequency by 1--2\%, which may have a significant impact on ground-layer turbulence. Further, we note that while the conditions themselves are not necessarily changing significantly, the combination of conditions that lead to dome closures (i.e.
freezing conditions, increased summit wind speeds and/or high humidities) are worsening to the point that the number of closure conditions have more than doubled in the last 20 years. Importantly, we find that the Fried parameter has not changed in the last 40 years, suggesting there has not been an increase in optical turbulence strength above the summit. Ultimately, more data and data sources---including profiling instruments---are needed at the site to ensure continued monitoring into the future and to detect changes in the summit climate.
}


\begin{document}

\maketitle

\section{Introduction}
With increasing global temperatures, weather around the world is changing \citep{ipcc_2019}. With more extreme weather events being attributed to climate change, research is focused on the impact of climate change on specific fields and industries. Astronomy may also be contributing to the crisis with the $CO_2$ emissions of an astronomer higher than the average adult (for example, 40\% higher in countries such as Australia)~\citep{nature_2020, Clery_2020}. The impact on climate of travel to conferences and meetings, operating observatories, and super-computing in turn could be impacting the quality of our astronomical sites and in the future limit ground-based astronomy. Recently, Cantalloube et al. (2021)~\cite{Cantalloube_2021} highlighted the need for in-depth study of the impact of climate change on astronomical observatories around the globe. Their recommendation was based on their investigation of parameters at the Paranal Observatory in Chile, where they found an increase in temperature and also surface layer turbulence. We investigate the climatology at Maunakea, one of the world's primary sites for ground-based astronomy, in order to determine if, and how, conditions may have evolved over the previous decades.

We specifically focus on the impact weather has on performing astronomical observations in the optical and near-infrared/infrared wavelengths at night. We first focus on the summit weather itself. Conditions for opening the telescope dome require that there be no precipitation, that the wind speed be below a specific threshold, and that the temperature and relative humidity present no risk of condensation forming on the primary telescope mirror (e.g., \cite{keckwebsite}). Should the nominal behavior of these parameters change, it could have a significant impact on the amount of time that observations can be made throughout a given year. 

It is also possible that the conditions, while not severe enough to prevent operating of the telescope, degrade the image quality to such an extent that the ability to further improve current high-resolution imaging becomes limited. For example, atmospheric turbulence can lead to the distortion of images through the introduction of wavefront aberrations caused by fluctuations in the index of refraction in air. Techniques---such as the use of adaptive optics (AO) and post-processing algorithms---are able to mitigate some of the distortion, however, they are limited in what they can remove and not all instruments can benefit from such corrections (e.g., not all instruments are AO-fed). Should turbulence be increasing at the site, it means that future telescopes will need to be designed with more high-order AO systems (to correct for higher spatial frequencies) that can also correct for more turbulence (i.e., large stroke of the deformable mirror to correct for greater optical path differences due to changes in index of refraction). This is particularly important when considering the building and operation of future extremely large telescopes such as the Thirty Meter Telescope (TMT); should the atmospheric turbulence be worsening, the performance requirements of future AO systems currently being designed might not be achieved. Recent work by Lee et al. (2019)~\cite{Lee_2019} shows that increasing wind shear above the North Atlantic will lead to significant increases in turbulence that could in turn impact air travel between North America and Europe. We investigate whether a similar increase in turbulence is also found above Maunakea (where a similar Jet Stream feature exists), and whether that in turn leads to an increasing optical wavefront error. An important factor when considering turbulence is the ``seeing'' (related to the Fried parameter, $r_0$) which is the full-width-half maximum (FWHM) of the imaged point-spread-function (PSF) without AO correction. We look specifically at the vertical structure function of the index of refraction, the $C_n^2$ profile, which is what ultimately determines $r_0$ and therefore the seeing. Larger values of $r_0$ at Maunakea have been shown to be correlated to higher wind speeds \citep{Chun_2009,Lyman_2020}. At the same time, slow wind speeds might increase the occurrence of the low-wind effect seen by the Subaru telescope on Maunakea \citep{Vievard_2019} which is due to slow moving wind within the dome not allowing for proper cooling. We therefore consider both trends in $r_0$ and $C_n^2$, as well as the summit wind speed. 

The rest of this paper is structured as follows. In Sects.~\ref{s:metdata} and \ref{s:turbdata}, we present an overview of the data and methods used in our analysis of the meteorological and turbulent characteristics of the summit, respectively. We then present the results of our analysis in Sect.~\ref{s:results}, beginning with a general overview of the sites' climatology over the previous decades, followed by an analysis of trends and impacts on observing/observable conditions. This is followed by a discussion of the results in Sect.~\ref{s:disc} and a summary of the conclusions in Sect.~\ref{s:concl}. 

\section{\label{s:metdata}Meteorological Data and Methods}

\subsection{Meteorological Data}

\begin{figure}
    \centering
    \includegraphics[width=0.5\textwidth]{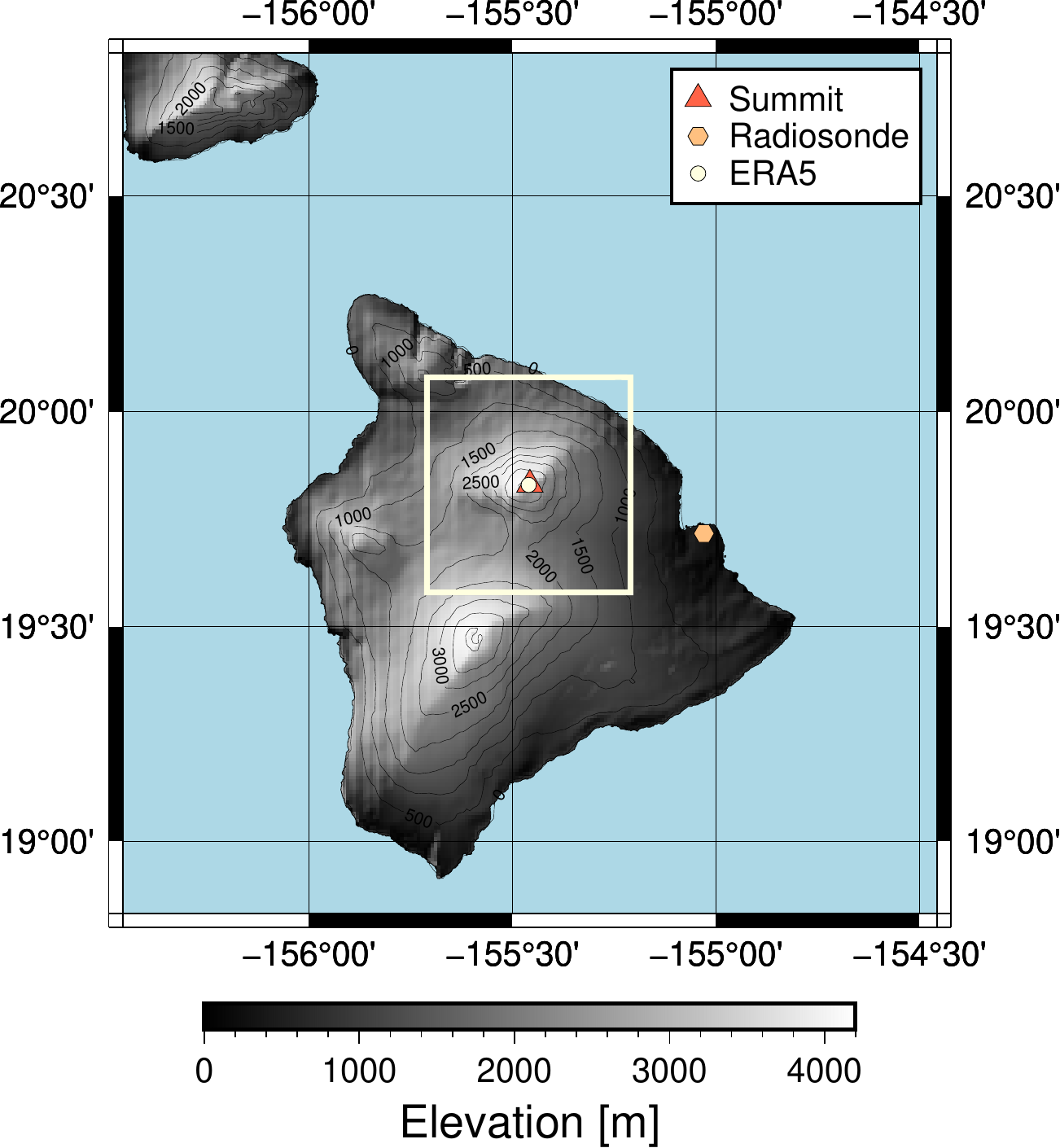}
    \hspace{0.5cm}\includegraphics[width=0.45\textwidth,trim=0cm -3cm 0cm 0cm]{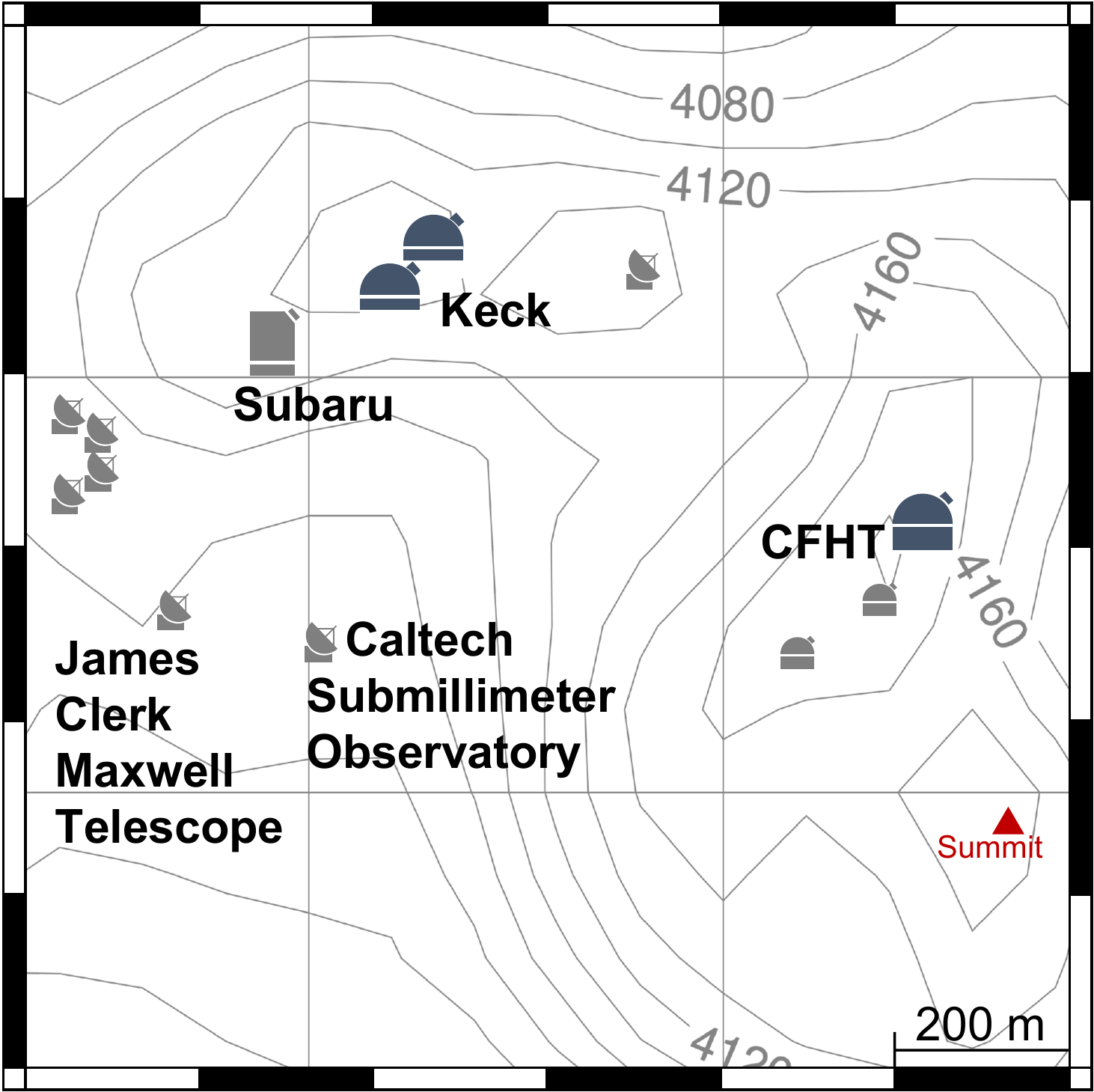}
    \caption{\label{f:datamap} Maps showing the geography of Maunakea and the locations of different data sources. Left: The position of Maunakea on the Island of Hawaii is indicated with the triangle. The radiosonde release point (Hilo) is shown with the orange hexagon. The ERA5 grid cell is indicated by the square, with the circle indicating the centre of the grid cell. The contours illustrate the topography of the islands, while the blue represents the ocean. Right: Sketch of Maunakea's summit showing some telescope locations on top of elevation contours (every 20~m). The METEO and MASS data are acquired from instruments located on a tower near CFHT. The basemaps were created using the PyGMT package in Python~\cite{pygmt_2022}.
    }
\end{figure}

\begin{figure}
    \centering
    \includegraphics[width=\textwidth]{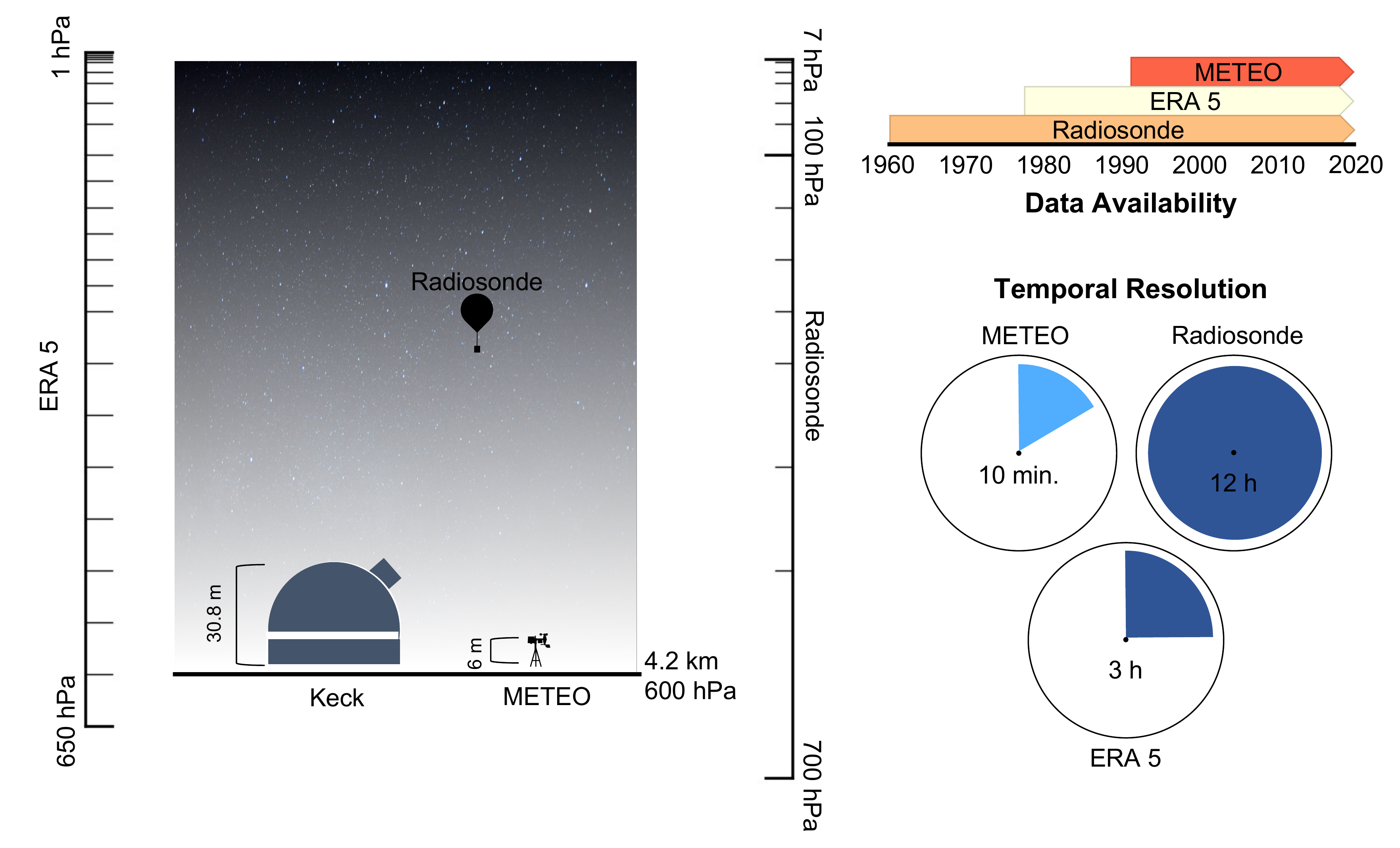}
    \caption{\label{f:datatime}Schematic illustrating the data characteristics. Left: vertical resolution with only the relevant levels at/above the summit. The ERA5 and radiosonde data are indicated in coordinates of atmospheric pressure. As noted in Sect.~\ref{s:rds}, the vertical resolution of the radiosonde data increased over time; as such, we show representative ranges in the years 1960 (up to 100~hPa) and 2020 (up to 7~hPa). The Keck dome is shown for illustration only. Upper right: availability of data in time. Lower right: temporal resolution of the data. Note the METEO data are available at different rates over the course of the 30 years; however, we resample them to a 10-minute average value.}
\end{figure}

We use three types of meteorological data in our analysis: in situ observations made at the summit, radiosonde profiles, and a numerical re-analysis. The left panel in Fig.~\ref{f:datamap} presents a map of the area around the Island of Hawaii (also referred to as ``The Big Island''), showing the location and extent of different data sources in relation to the summit of Maunakea, while the right panel includes the layout of the summit with major telescopes shown. Figure~\ref{f:datatime} further illustrates the vertical and temporal resolution of the various datasets. We deliberately sought data that are available for roughly 30 years or more in order to be able to extract meaningful climatologies. The individual data sources are described in the following subsections.

\subsubsection{In Situ Meteorological Data}
In situ meteorological data is from the Canada-France-Hawaii Telescope (CFHT) meteorological tower located at the summit of Maunakea. These data---referred to as the METEO (meteorological) data---are available from 1991 to present day, and can be downloaded from the Maunakea Weather centre\footnote{\url{http://mkwc.ifa.hawaii.edu/archive/}}. Included in the dataset are 1-minute observations of wind speed ($U$), temperature ($T$), atmospheric pressure ($p$), and relative humidity (\emph{RH}).

We use the METEO data in order to provide an indication of the weather on top of the mountain at the location of the telescopes, as well as a ``ground truth'' benchmark for assessing the quality of the other data. Unlike the other data sources, the METEO data do not provide any vertical resolution. In order to smooth noisy data, we average the 1-minute values over 10-minute periods.

\subsubsection{\label{s:rds}Radiosonde}
Radiosondes are instrument platforms---typically carried by balloons---that are used to profile the properties of the atmosphere from the surface through the stratosphere~\citep{wmo14}. Observations are made as the balloon ascends, with values recorded at mandatory vertical levels (that change throughout the record), or at significant (thermo-) dynamic locations in the profile~\citep{schwartz92}. A typical ascent lasts roughly two hours; since the balloons are not steered, this means they can potentially drift up to hundreds of kilometers from their release point (e.g., \cite{Seidel_2011, Laroche_2013}). Yet, work by Bely (1987) \citep{Bely_1987} demonstrates good agreement between in situ and radiosonde observations. In Sect.~\ref{s:validn}, we further compare the radiosonde measurements to summit observations in order to determine whether or not the results can be compared with confidence for Maunakea. 

Radiosonde observations have been made on Hawaii since the 1950s. With improvements to instruments, the number of levels at which an observation is made has increased from roughly 10 above-summit locations to greater than 100 in recent years (Fig.\ref{f:datacomp}). Released twice per day at the Hilo International Airport (Fig.~\ref{f:datamap}), the radiosonde data provide a useful secondary verification for the re-analysis data set we use (ERA5). They also provide quasi-in situ vertical information, which the METEO data is unable to provide. From the radiosonde, we have vertical profiles of the temperature, humidity, and wind speed. The radiosonde data were downloaded from the NOAA/ESRL Radiosonde Database\footnote{\url{https://ruc.noaa.gov/raobs/}}. In the following, we refer to the radiosonde observations by the abbreviation, RDS.

\subsubsection{ERA5 Re-Analysis}
Re-analysis datasets are constructed by running \emph{a posteriori} simulations of numerical weather models and assimilating available in situ and remote-sensed data (e.g.,  radiosonde, weather station, surface temperature data) in order to provide an historical estimate of conditions over the globe. In essence, they provide time-space interpolations of sparse data. The European Centre for Medium-Range Weather Forecasts (ECMWF) produces its ERA-5 re-analysis with 137 vertical levels (roughly 68 above Maunakea's summit); the data are available for download as 3-hourly mean values from the Copernicus Climate Data Store\footnote{\url{https://climate.copernicus.eu/climate-reanalysis}} on a horizontal grid of $0.25^\circ \times 25^\circ$ and interpolated to 25 vertical levels above the summit (650--1~hPa). As with the in situ observations, the reanalysis includes values of temperature, wind speed, atmospheric pressure, and at least one metric of humidity from which other metrics can be determined. We downloaded the data at the closest model grid cell to the summit location of Maunakea, which is centred almost directly at the summit.

\subsubsection{\label{s:validn}Validation with In Situ Observations}
\begin{figure}
    \centering
    \includegraphics[width=\textwidth]{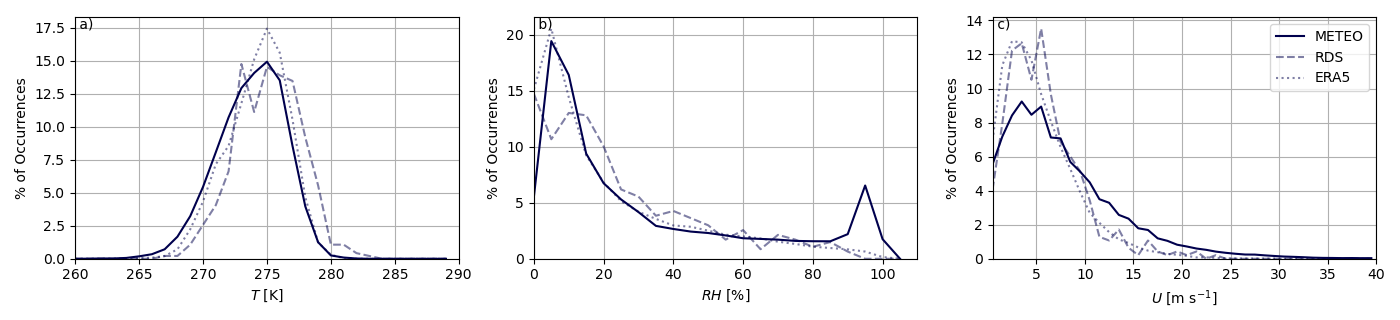}
    \caption{\label{f:datacomp}Histograms comparing summit meteorological conditions at Maunakea from different data sources. In the case of the ERA5 and RDS data, we take the value closest to the summit altitude of ~4.2 km above sea level. a) Temperature. b) Relative humidity. c) Wind speed.}
\end{figure}

We present a very brief comparison of the statistics of the meteorological data used in our analysis in this section. We do this as a validation step to ensure that our data---particularly the reanalysis data---are reasonably representative of the observed conditions. While exact instantaneous values may differ, the statistics (e.g., mean and variance) of the different data sets should be similar in order to facilitate comparison between datasets. A series of histograms is presented in Fig.~\ref{f:datacomp} illustrating the distribution of summit-level temperature, humidity, and wind speed. Given that the radiosonde and ERA5 reanalysis data are reported on pressure levels, their measurements are not guaranteed to correspond precisely to the summit altitude of 4.2~km above sea level. As such, we select the observations taken at the level closest to the summit.

The METEO temperature distribution is broader than the others, with a slightly cooler peak. At the same time, while the ERA5 and in-situ METEO data have nearly identical distributions in the range from roughly 10\%--70\%, the in situ METEO contains much higher relative humidity values, approaching saturation almost 5\% of the time, while the other sources almost never reach saturation. This discrepancy could be related to local topographic effects (e.g., upslope advection of air leading to saturation) that are not resolved by the vertical profiling of the free atmoshere. The greatest deviation occurs in the wind speed, where the METEO reaches wind speeds greater than 10~m~s$^{-1}$ far more often than in the radiosonde and reanalyses (which all agree with each other). This may also be a local effect due to summit topography, as discussed by Bely (1987)\cite{Bely_1987}.  While Bely corrects for deviations in wind between radiosonde and in situ observations, we are interested in the relative change on an instrument-basis and so do not make any adjustments to the reported observations. Some of the differences in distributions may be due to the much higher rate of sampling of the METEO (here, 10-minute averaged vs $\ge$~3 hours); however, resampling the METEO data (e.g., taking a 3-h mean), does not bring the distributions closer (not shown). The fact that the vertical profiles are not sampling exactly at the summit may also lead to some of the differences seen. 

Ultimately, as a result of this brief comparison, we conclude that the differences in the underlying summit distributions will potentially lead to significant differences when determining in situ conditions for observing. As such, we only use the in situ METEO data for analysis of summit conditions. The ERA5 reanalysis does, however, do a good job of reproducing the observed properties of the radiosonde (both near the summit as shown here, and vertically; not shown). We can therefore use the ERA5 profiles for the the turbulent parameter estimation (see Sect.~\ref{s:turbdata}) for which summit values alone are insufficient. Using the ERA5 reanalysis rather than the radiosonde profiles allows for a much higher temporal resolution on a consistent vertical grid.

\subsection{Meteorological Methods}
In order to ensure that we are looking at nocturnal conditions, we use a strict window, limiting our analyses to the data taken at times between 2100 and 0600 local Hawaii Standard Time (HST; UTC+10). This means we have one radiosonde profile and four ERA5 values per night.

Meteorology influences observational astronomy in two primary ways. 1) The meteorological properties affect the quality of the observations through changes in the index of refraction and turbulent properties of the atmosphere, as well as (for longer wavelengths of observation) the background emissivity of the atmosphere. 2) Whether or not observations can even take place is also dependent on the weather. A simple example is risk of condensation; water condensing on the telescope not only reduces the transfer of light throughout the system but can also damage the surface of mirrors. Many telescopes have specific operating conditions where the dome cannot be opened (or must be closed) if conditions reach certain thresholds. In our investigations, we are therefore concerned with understanding the overall trend in the meteorological variables, as well as any changes in observable conditions.

We perform our meterological analyses by binning the data into seasons classified according to three months as follows: spring (March, April, and May; MAM), summer (June, July, and August; JJA), autumn (September, October, and November; SON), and winter (December, January, and February; DJF). We also look at annual values. In order to better ensure reliable results, we restrict ourselves to periods where at least 80\% of the data are valid (i.e., no missing or invalid data).

We determine the long-term trends in meteorological conditions themselves based on the mean over the seasonal/annual period. We are interested in both the long-term trends at the summit, as well as in the column of atmosphere above the summit. We only consider the in-situ meteorological data to determine summit trends, while the reanalyses and radiosondes are used to provide an indication of the above-summit conditions.

Determining a season's (or year's) potential for observation requires the further step of first comparing the in-situ observations with the meteorological thresholds. At each observation time, we compare the available data to the threshold. If the data do not exceed the threshold, then the timestep is considered ``observable''. If, however, the threshold is exceeded, then it would be deemed ``unobservble''. Our thresholds are based on those listed by the Keck Observatory~\cite{keckwebsite} which provides guidelines for Observing assistants on when to close the dome.  The actual operation of the telescope will, of course, depend on the experienced decision making of the Keck personnel at the summit and so these thresholds are not absolute. They do, however, suffice for the purposes of our analysis to give an indication of operational feasibility. Briefly, the primary thresholds for observable conditions that we consider here are: $U < 20$~m~s$^{-1}$, \emph{RH}~$< 95$\%, and $T-T_{dew} > 2$~K.  A more detailed overview of the observing criteria can be found at \cite{keckwebsite}.

\subsubsection{Precipitable Water Vapour}
The total precipitable water vapour (\textit{PWV}) is the total amount of water within a column of the atmosphere:
\begin{equation}
    \textrm{\textit{PWV}}=\frac{1}{\rho_wg}\int_{P_1}^{0}q(P)dP
\end{equation}
where $\rho_w$ is the density of water vapour, $g$ the acceleration due to gravity, $q$ specific humidity, and $P$ the atmospheric pressure. \textit{PWV} is an important parameter for observing in the (near-) infrared (NIR/IR) and submillimeter wavelengths as water radiates at these wavelengths dominating background radiation. Water also introduces phase aberrations for longer wavelengths as it becomes a source of fluctuations in the index of refraction (the fluctuations are driven by temperature for optical/NIR wavelengths)~\cite{Colavita_2004}. With large amounts of water present in the atmosphere it becomes difficult to observe in these wavelengths from the ground. \textit{PWV} values around 5-10~mm can render it impossible to make scientifically impactful observations in the K-band (central wavelength of 2.2 $\mu$m) for science cases such as the direct imaging of exoplanets. For sub-millimeter, certain bands can only be observed when the \textit{PWV} is less than 1~mm. We use the James Clerk Maxwell Telescope, JCMT, weather bands\footnote{\label{note:JCMT}\url{https://www.eaobservatory.org/jcmt/observing/weather-bands/}} to bin the calculated \textit{PWV} and study the long term behaviour \textit{PWV} for these wavelengths.




\section{\label{s:turbdata}Turbulence Data and Methods}

\subsection{MASS-DIMM data }
The Canada France Hawaii Telescope (CFHT; Fig.~\ref{f:datamap}) provides nightly time series of the total seeing (i.e., $r_0$) and vertical profiles of the index of refraction structure function, $C_n^2$, estimated by its Differential Image Motion Monitor (DIMM) and Multi-Aperture Scintillation Sensor (MASS). The MASS $C_n^2$ profiles are estimated for fixed altitudes of 0.5, 1, 2, 4, 8, 16 km above the telescope and are made approximately every 2 minutes. The data are available from 2009 to present. The MASS instrument has limited ability to measure the turbulence accurately for the first altitude of 0.5 km as it is blind to some of the turbulence in the layer and can only measure the seeing in the free atmosphere. The DIMM, however, measures the integrated turbulence for the entire column allowing $r_0$ to be estimated. By combining the DIMM/MASS, an estimation of the ground-layer turbulence can be made.

\subsection{\label{sec:turb_method}Estimating Turbulence Parameters}
In ground-based optical and NIR astronomy, a few key parameters are used to describe atmospheric turbulence in a way that is meaningful for observing, including: the $C_n^2$, the Fried parameter ($r_0$), and the atmospheric coherence time ($\tau _0$). These parameters are either directly related to the image quality or have meaning for the performance of an AO system. $C_n^2$ is the structure function of the index of refraction as a function of altitude. At the observed wavelengths, fluctuations in the index of refraction cause the optical path differences (phase errors) that limit image quality and resolution of larger telescopes. The Fried parameter is related to the integrated $C_n^2$ and describes the total impact of the atmosphere. With units of length, $r_0$ can also be estimated as an angular separation in units of arcseconds giving the more commonly used value of "seeing" that astronomers report as it relates to the FWHM of an aberrated PSF.  Finally, $\tau _0$ describes how quickly the turbulence is changing above the telescope. It is related to the wind speed and the turbulence strength profile. With these parameters we can predict the quality of observed data and the achievable image resolution. In this section we outline the calculation of these values. 
\subsubsection{Determining Structure Function of the Index of Refraction from Re-analysis Data}
The ERA5 re-analysis data contain temperature and wind values at 25 pressure levels above the Maunakea summit (Fig.~\ref{f:datatime}). This corresponds to a value every few kilometers which is relatively coarse, though much finer resolution than MASS data. We match synthetic $C_n^2$ profiles generated from ERA5 to the coarse profiles as measured at CFHT. Here we do not aim for exact instanaeous matches rather estimate the mean $C_n^2$ in order to calibrate out ERA5-derived profiles. This provides us with rough estimates of the local $C_n^2$ (although with such coarse resolution it cannot be a 'local' estimate). Since the reanalysis has more layers than CFHT, we re-sample the $C_n^2$ data by summing the local $C_n^2 dh$. This provides us with the $C_n^2$ in $m^\frac{-1}{3}$, allowing us to compare to the CFHT measurements. Below we outline our methodology for calculating the $C_n^2$ values from ERA5 data.

Using same methodology as \citet{Osborn_2018}, we use the modified Gladstone's relationship~\cite{Masciadri_2016} to write $C_n^2$ as a function of the temperature structure function, $C_T^2$.
\begin{equation}
  C_n^2 =(80*10^{-6} P/T\theta)^2 C_T^2  
\end{equation}
where $\theta$ is the potential temperature,

\begin{equation}
    \theta = T \frac{P_0}{P}^{\frac{R}{c_P}}, 
\end{equation}
and $P_0 =1000 mbar$ and $\frac{R}{c_p} =0.289$. 

From Tatarskii et al. (1971)~\citep{Tatarskii_1971}, $C_T^2$ as a function of altitude ($z$) is estimated using the potential temperature gradient, and the scale of the largest energy scale of the turbulent flow, $L$.
\begin{equation}
    C_T^2= kL(z)^{4/3} \left(\frac{\delta \bar{\theta} (z) }{\delta z}\right)^2
\end{equation}
\begin{equation}
    L(z)=\sqrt{\frac{2E}{\frac{g}{\theta(z)}\frac{\delta \bar{\theta} (z) }{\delta z}}}
\end{equation}
$k$ is an unknown dimensionless constant that is calibrated against $C_n^2$ data; in reality it encodes information about the stability of the atmosphere. In Osborn et al. (2018)~\cite{Osborn_2018}, the authors found a $k$ value of 6 for a global calibration. However, $k$ can be determined for not only a specific site but also be altitude dependent. In this work we calibrate $k$ using approximately 10 years of MASS data starting from 2011 and find a value for each altitude: 6.3, 10.3, 25.6, 11.8, 18.2, and 12.0 going from the lower to higher altitudes, respectively. $E$,~the turbulent kinetic energy, is given by the square of vertical wind shear as done in Osborn et al. (2018)~\cite{Osborn_2018}:
\begin{equation}
    E= \left(\frac{\delta u }{\delta z}\right)^2 +\left(\frac{\delta v }{\delta z}\right)^2 . 
\end{equation}
\subsubsection{Determining Fried Parameter and Atmospheric Coherence Time}
From Hardy (1998)~\cite{Hardy_1998}, $r_0$ for light at 500~nm in the zenith direction can be calculated from the $C_n^2$ profile
\begin{equation}
    r_0 = 0.423\left ( \frac{2 \pi}{\lambda}\right )^2\left(\int C_T^2 (h) dh \right)^{-3/5} 
\end{equation}

From the Fried parameter, $r_0$ and the effective wind speed, $V_eff$, we can calculate the coherence time, $\tau_0$ following Hardy (1998)~\cite{Hardy_1998}.  

\begin{equation}
    \tau_0 = 0.314 \frac{r_0}{V_{eff}}
\end{equation}

with

\begin{equation}
    V_{eff}= \left[\frac{\int_{0}^{\infty}C_n^2(h) U(h)^{5/3} dh}{\int_{0}^{\infty}C_n^2(h) dh}\right]^{3/5}
\end{equation}.
\section{\label{s:results}Results}
\subsection{Meteorology}
\subsubsection{Climate at Maunakea}\label{s:climate_mk}
\begin{figure}
    \centering
    \includegraphics[width=\textwidth]{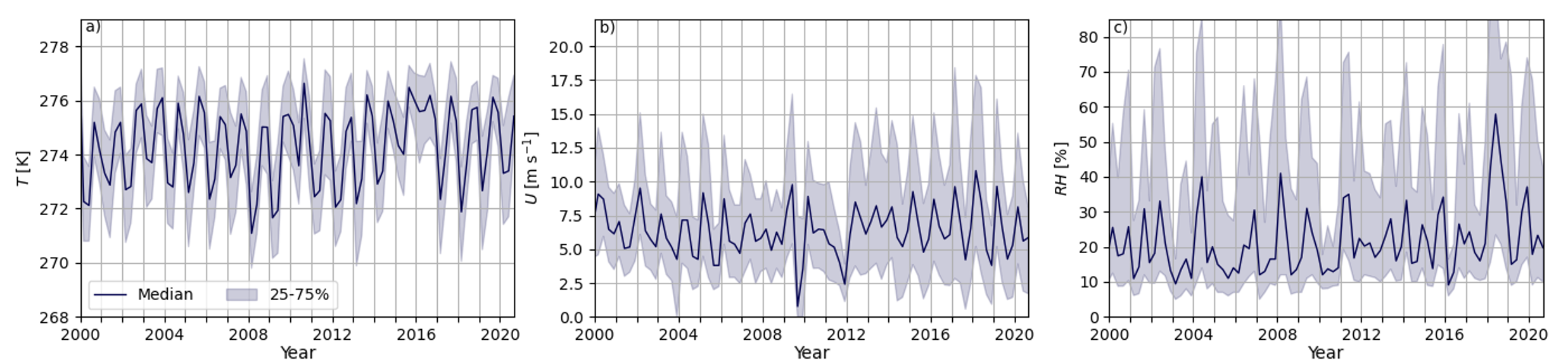}
    \caption{\label{f:seasonalvariance}Median of the seasonally binned nocturnal METEO observations, with shading indicating the range spanned between the 25$^\textrm{th}$ and 75$^\textrm{th}$ percentiles. a) temperature, b) wind speed and  c) relative humidity.}
\end{figure}

\begin{figure}
    \centering
    \includegraphics[width=0.75\textwidth]{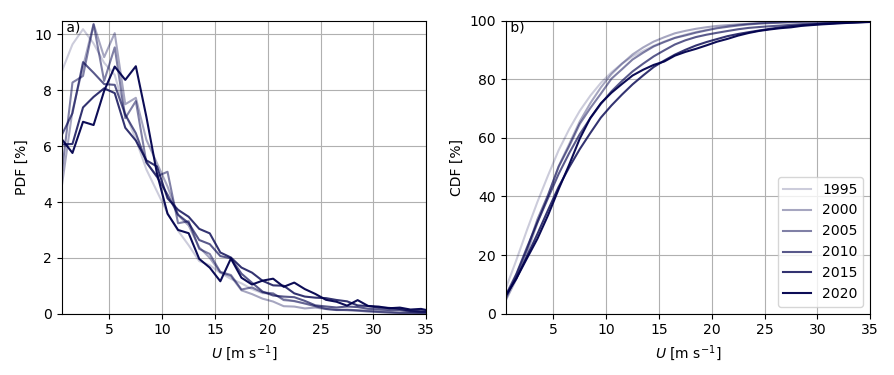}
    \caption{\label{f:windchange}5-year binned values of wind speed shown as a) probability density and b) cumulative density functions.}
\end{figure}

\begin{figure}
    \centering
    \includegraphics[width=0.57\textwidth]{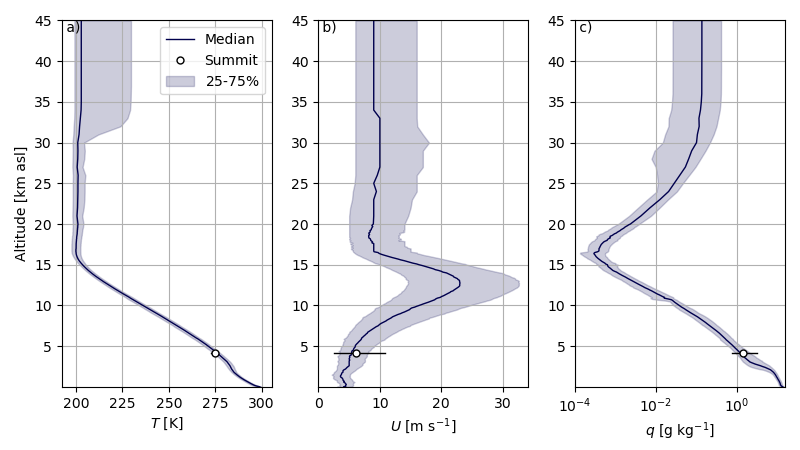}
    \caption{\label{f:meanrds}Nocturnal profiles of a) temperature b) wind speed, and c) specific humidity as measured by the radiosonde. The lines indicate the median value over 40 years  with the shading indicating the range spanned between the 25$^\textrm{th}$ and 75$^\textrm{th}$ percentiles. The circle and whiskers represent the median and percentile range of the equivalent METEO observations.}
\end{figure}

We first present an overview of Maunakea's climate based on the assembled data, in order to provide a context for subsequent analysis. Figure~\ref{f:seasonalvariance} shows the seasonal-median summit values of temperature, wind speed, and relative humidity from the year 2000 to present as recorded at the CFHT weather station. All three parameters demonstrate a degree of seasonal variability though overall relatively stable characteristics. Temperature is within a few degrees of freezing throughout the year and summit wind speeds are typically around 5--7~m~s$^{-1}$. In general, the atmosphere is dry with relative humidity around 20\%, but with significant variability in the record (illustrated by the shading), including an increase in the standard deviation with time.
Of the standard meteorological variables, wind speed is the only parameter with any statistically significant increase in median value at the summit (Fig.~\ref{f:windchange}) with 5-year averaged speeds increasing over 30 years. For example, there is a rightward shift in the peak of the wind-speed distribution, with speeds above 15~m~s$^{-1}$ increasing by 1--2\%. Overall this does not have a significant impact on the mean summit windspeed, but it does indicate a greater likelihood for increased ground-layer turbulence and, by extension, wind buffeting as the wind interacts with the dome structures themselves~\cite{tmt_wind_buffet}.
The median vertical structure of the atmosphere is shown in Fig.~\ref{f:meanrds}. The jet stream layer is visible as a maximum in wind speed between 10--15~km above sea level, with maximum wind shear at the top and bottom of the layer. Within this layer, the wind predominantly blows from west to east, with equal likelihood of small northerly or southerly deviations (not shown). In general, temperature decreases near-adiabatically with height up to roughly the top of the jet stream where it becomes roughly constant. Specific humidity also decreases within the troposphere, before increasing above 15~km.

\begin{figure}
    \centering
    \includegraphics[width=\textwidth]{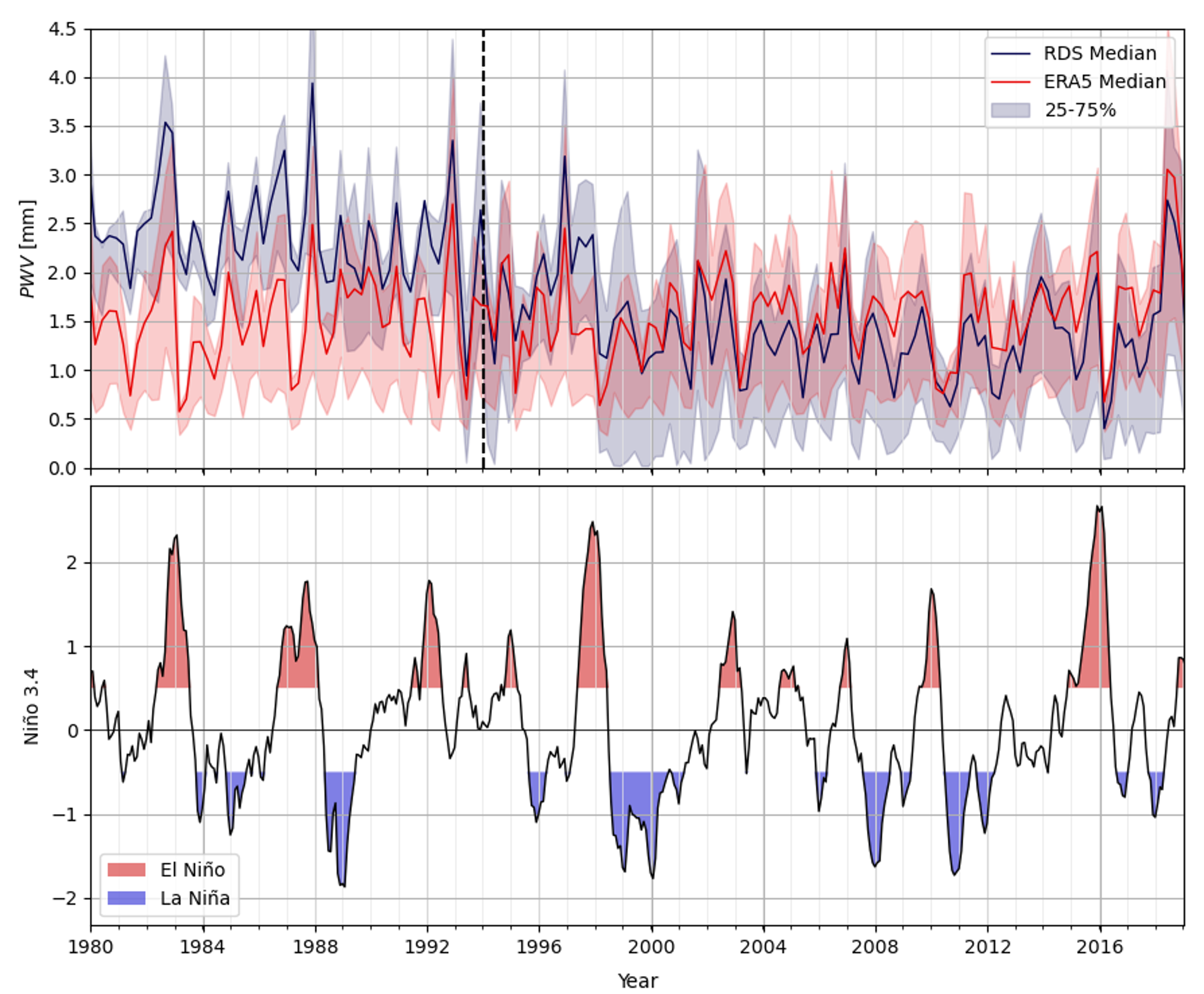}
    \caption{\label{f:pwvenso}Seasonal \textit{PWV} compared with ENSO conditions.}
\end{figure}

The precipitable water vapour, related to specific humidity, shows considerable variability over the period from 1980 to present (Fig.~\ref{f:pwvenso}). \textit{PWV} varies between 0.5--3~mm in both the radiosonde- (RDS) and ERA5-calculated values, with no significant long-term trend in the seasonal median. We also compare the \textit{PWV} to the El Ni\~na-Southern Oscillation (ENSO) in Fig.~\ref{f:pwvenso}. Minima in the signal appear to follow peaks in El Ni\~no and subsequent transitions to La Ni\~na conditions, providing a first-order predictor of conditions. Interestingly, there is a point before 1995 where the ERA5 and RDS values disagree, after which they are aligned, although the RDS does consistently lead to lower minima than the ERA5 estimates. The seemingly abrupt change in RDS values is likely due to an increase in the number of vertical levels sampled by the radiosonde between roughly 1995 (dashed line) and 1998, rather than any relevant climatological factor.

\begin{figure}
    \centering
    \includegraphics[width=\textwidth]{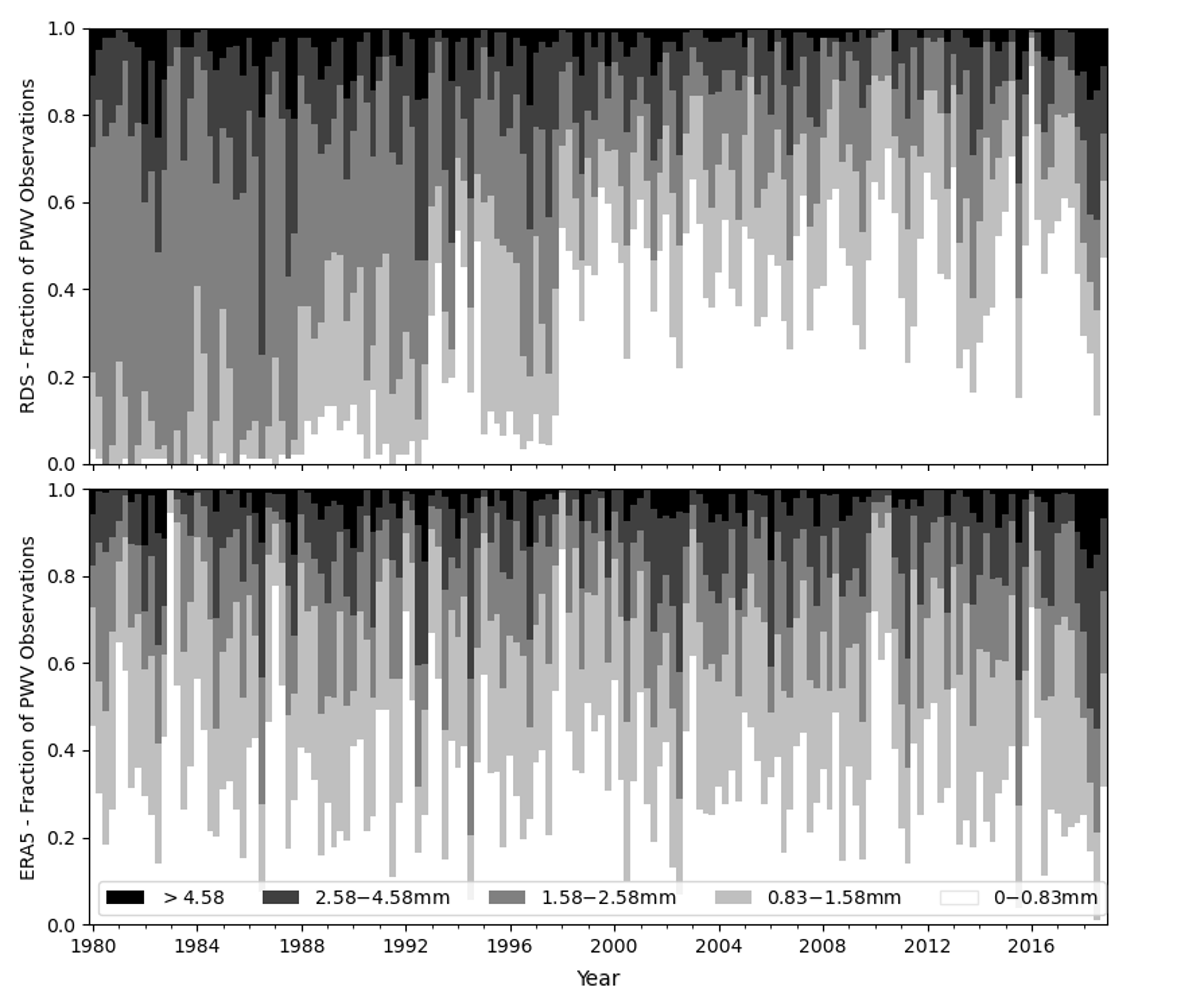}
    \caption{\label{f:seaspwv}Fraction of seasonal total atmospheric \textit{PWV} in the JCMT's weather bands\textsuperscript{\ref{note:JCMT}}.}
\end{figure}

Figure~\ref{f:seaspwv} further highlights the seasonal distribution of total atmospheric \textit{PWV} in different bins. The RDS observations indicate that, post-90's, \textit{PWV} is less than 0.83~mm almost 50\% of the time, providing optimal conditions. The ERA5 reanalysis estimates a much lower fraction, though with \textit{PWV}~$<$~1.58~mm at least 50\% of the time. The difference in estimates may be due to the different vertical resolution of the profiles. After the mid-90's, the RDS data have a much higher vertical resolution than the ERA5 profiles. The consistent heights of the reanalysis data, however, allow for a long-term comparison of PWV to be made. While there are obvious variations over the previous 40 years, they are of a cyclical nature with no obvious trend.

\subsubsection{Weather-Related Dome Closures}
\begin{figure}
    \centering
    \includegraphics[width=\textwidth,trim=0cm 0cm 0cm 1cm,clip]{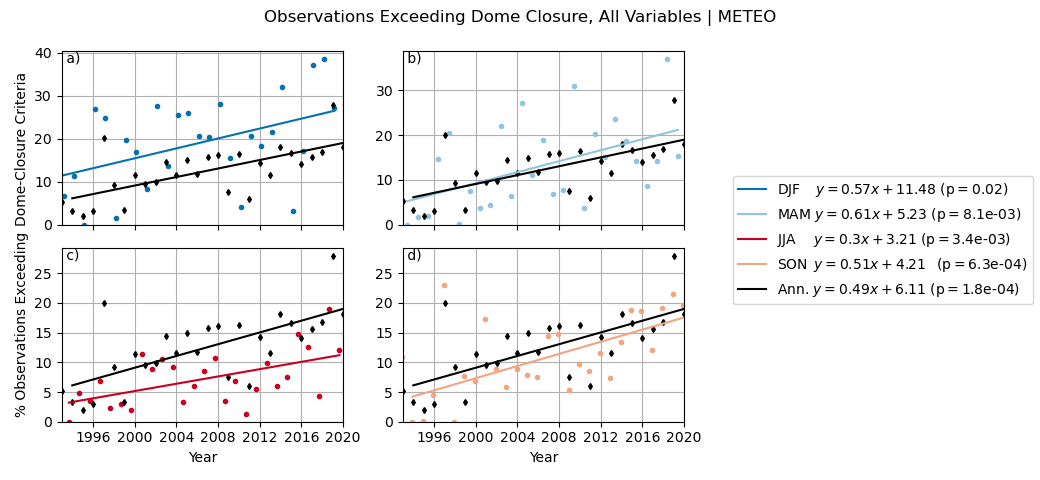}
    \caption{\label{f:threshobs}Percentage of all nighttime observations where the METEO-observed conditions exceed the dome-closure criteria for the Keck Observatory. The dots show the seasonal mean, while the lines indicate the linear regression. The colours distinguish seasonal values, while the black line (same in each panel) indicates the annual trend. The regressions and their significance (p-value) are shown in the legend. a) Winter. b) Spring. c) Summer. d) Autumn.}
\end{figure}

We next investigate the meteorological conditions that can lead to dome closures at the summit, using the Keck values as a guide. Figure~\ref{f:threshobs} plots the fraction of all METEO observations that exceed the dome closure criteria (i.e., the total number of times the criteria are exceeded, divided by the total number of observations made) This is analogous to the amount of time that the dome would need to be closed in a season, though not directly comparable due to operational considerations such as the waiting time needed before the dome can be re-opened. We plot seasonal (different panels) and annual values (black dots in each panel). In all cases, there is a significant worsening trend in the fraction of observations that exceed the criteria, with an annual increase of 0.49\% per year. The greatest increase is seen in spring (0.61\% per year) and the lowest in summer (0.3\% per year). These trends equate to a  near-tripling in fraction of conditions requiring dome closure over the 30-year period. The trends are driven primarily by increasing summit winds in winter, spring, and autumn, while increasing humidity drives the trend in the summer (not shown).

\begin{figure}
    \centering
    \includegraphics[width=\textwidth,trim=0cm 0cm 0cm 1cm,clip]{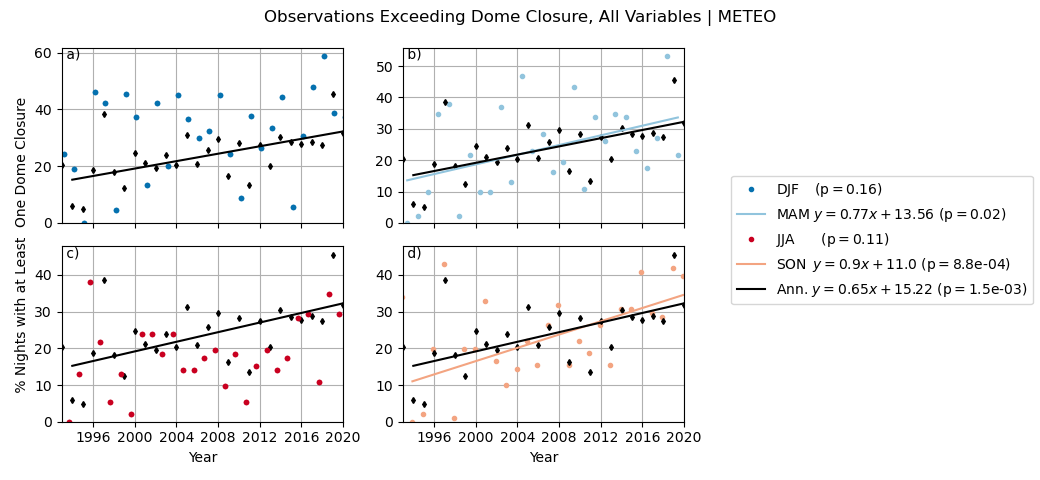}
    \caption{\label{f:threshnights}Same as Fig.~\ref{f:threshobs}, but now showing the percentage of unique nights on which at least one METEO observation exceeds the dome-closure criteria for the Keck Observatory.}
\end{figure}

While the total fraction of observations is increasing, Fig.~\ref{f:threshnights} shows that this does not necessarily equate to the same change in unique nights impacted by weather (i.e., nights were at least one criterion is exceeded at least once in the night). Winter has the most nights affected by bad weather, but only spring and autumn show significant trends, leading to an annually significant trend of around 0.65\% of nights per year. Over the 30 years, however, this does mean a roughly doubling in nights impacted by bad weather, going from approximately 15\% to over 30\% of unique nights by 2020.

\subsection{Turbulence}\label{s:results_turb}
Next we look at the behavior of turbulence above the telescope that drives the changes in optical path difference that causes image distortion and limits resolution. We split the analysis of turbulence into two components: 1) the free atmosphere (starting from approx 0.5 km above summit) and 2) ground layer using $C_n^2$, $r_0$, and $\tau_0$. 
 \subsubsection{Turbulence in the free atmosphere}
\begin{figure}
    \centering
    \includegraphics[width=0.65\textwidth,trim=0cm 0cm 0cm 0cm,clip]{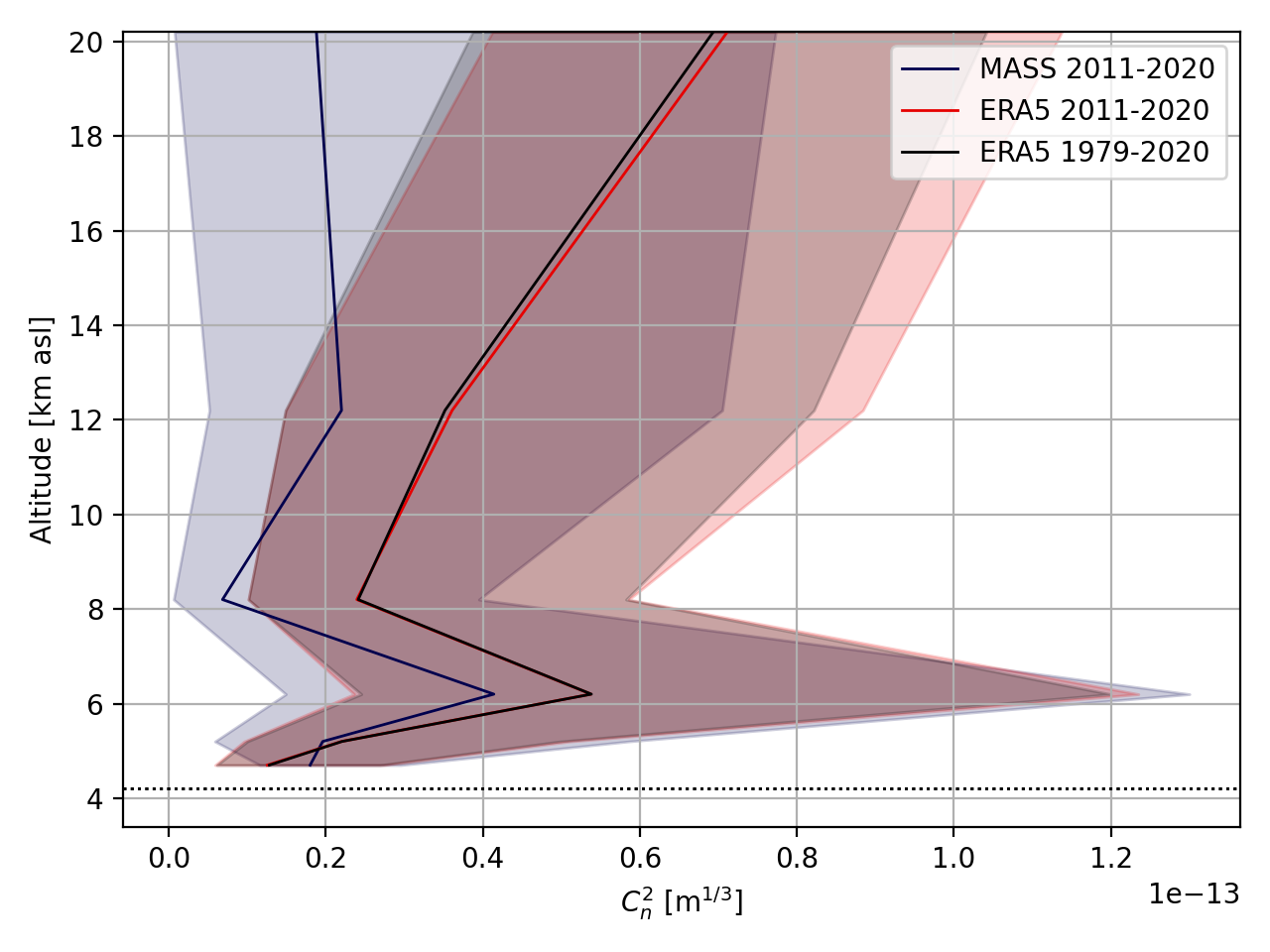}
    \caption{Free-atmosphere MASS and ERA5-estimated $C_n^2$ profiles. The solid line is the median profile and the shading the 25$^\textrm{th}$--75$^\textrm{th}$ percentile range. The ERA5 profiles are shown for the overlapping time period with the MASS (2011-2020) an the value over the entire dataset starting in 1979.}
    \label{fig:ERA5_CN2}
\end{figure}

\begin{figure}
    \centering
    \includegraphics[width=\textwidth,trim=0cm 0cm 0cm 0cm,clip]{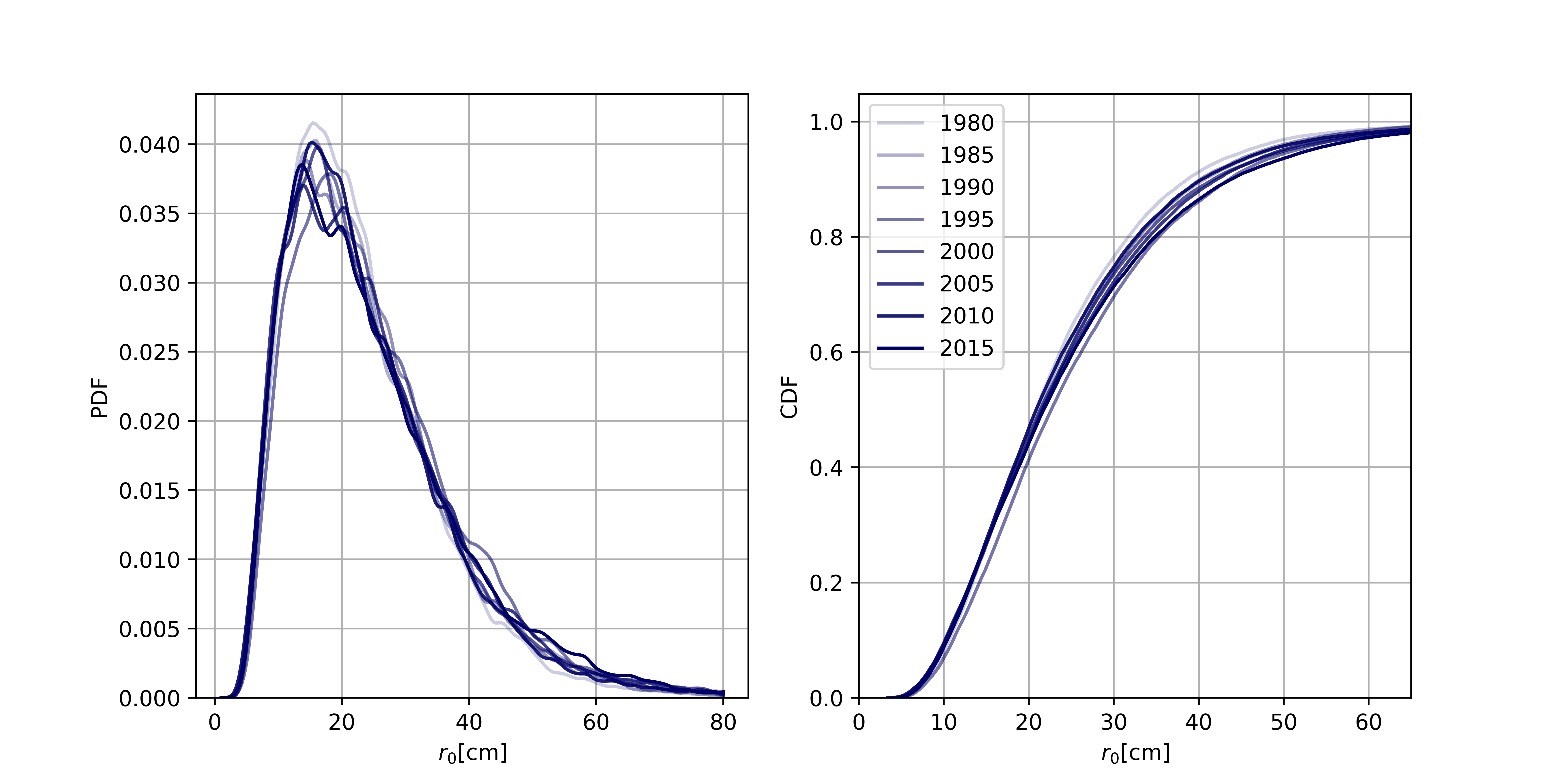}
    \caption{The probability density function (PDF; left) and cumulative density function (CDF; right) for 5-year-binned values of ERA5-derived values of $r_0$ of the free atmosphere. The lines go from lightest to darkest in chronological order with the legend indicating the start of the bin.}
        \label{fig:ERA5_r0_time}
\end{figure}

 From the equations outlined in Sec.~\ref{sec:turb_method}, the mean $C_n^2$ profile for the free atmosphere was calculated using the ERA5 reanalysis, re-sampled to the MASS/DIMM altitudes and then calibrated on on the overlapping data from 2011 to early 2020 by calculating the ratio of the mean profiles in time. The calibration was then applied to all the ERA5 profiles. We plot the results in Fig.~\ref{fig:ERA5_CN2}, showing the median instead of the mean in order to highlight differences in the profiles. Note this means that while the mean profiles are same due to the calibration, the extrema are different. As expected we have good agreement with the overlapping data, while all of the ERA5 data has a slighter smaller mean value than the most recent data, suggesting that the mean profile has increased in strength though within the error bars of the most recent profile. We calculate $r_0$ for the free atmosphere using the $C_n^2$ profiles, with 5-year-binned statistics of $r_0$ in Fig~\ref{fig:ERA5_r0_time}. From the PDF and CDF we see temporal variability in $r_0$ but no consistent trend toward better or worse values. These results suggest that the strength of turbulence has not changed in the last 40 years. 

\begin{figure}
    \centering
    \includegraphics[width=\textwidth,trim=0cm 0cm 0cm 0cm,clip]{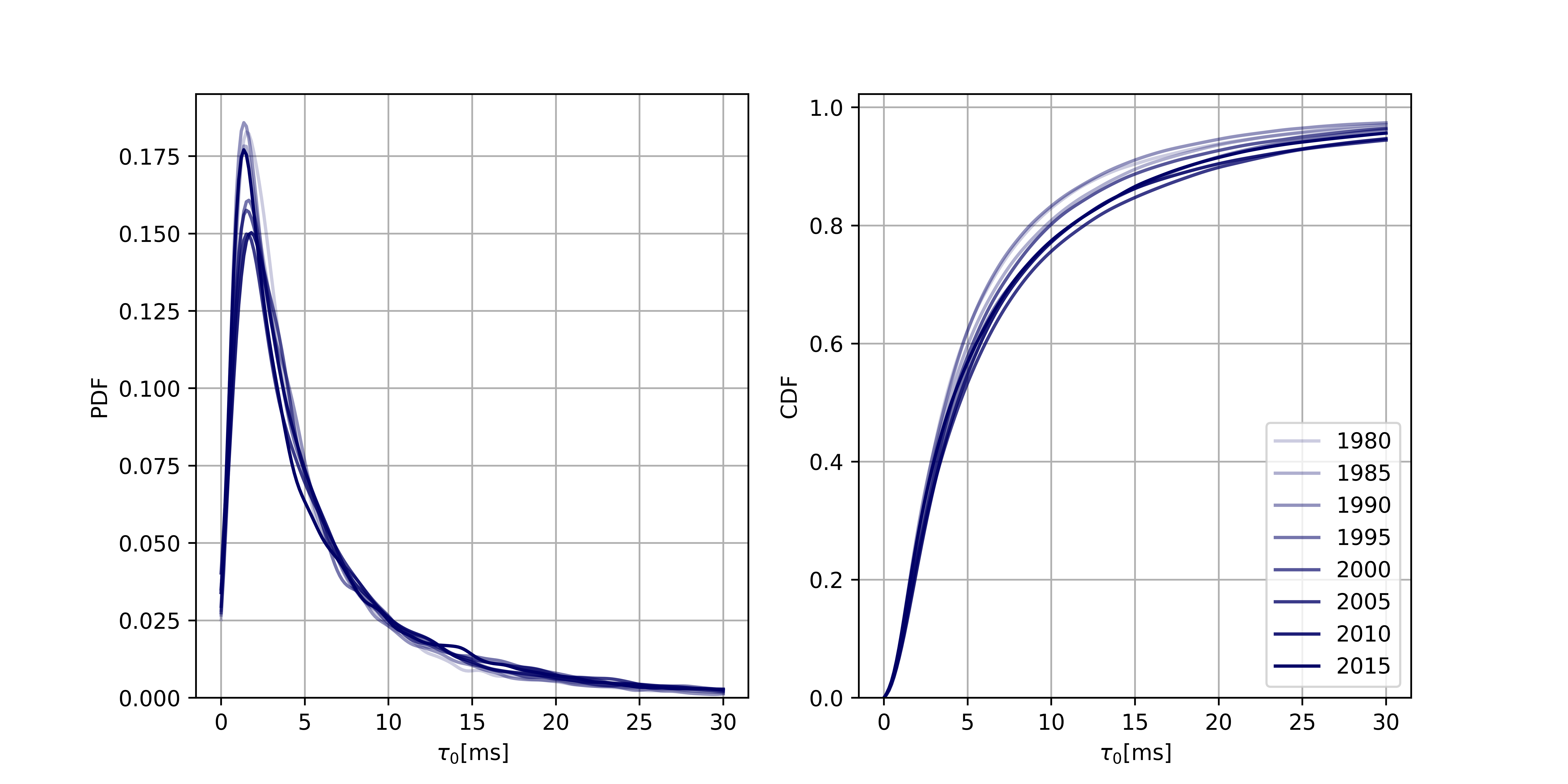}
    \caption{The same as Fig~\ref{fig:ERA5_r0_time} but for $\tau_0$ for the free atmosphere.}
        \label{fig:ERA5_tau_time}
\end{figure}

While the strength of the turbulence has remained constant, we look to the free atmosphere $\tau_0$ which can have a significant impact on the image quality for astronomical observations especially AO assisted imaging. Taking the effective wind speed, we calculate $\tau_0$ with Fig.~\ref{fig:ERA5_tau_time} showing the statistics of $\tau_0$ over the same temporal bins. From 1980-2014, the peak of the PDF is decreasing with time and from the CDF we see that the curves shift to the right indicating that the $\tau_0$ has more larger values between. There is, however, an increase in the peak of the PDF for 2015-2020. Looking more closely at the temporal sampling of the data, we are unable to confirm whether the changes are significant as the amount of data available within each bin varies by tens of percentages as well as the distribution throughout the year. Qualitatively, we see no evidence of changes to the wind speed in the free atmosphere at the MASS altitudes that we use for the $\tau_0$ calculations. 
 The free atmosphere behaviour, however, does not provide the complete story; we must also look to what the ground layer turbulence is doing. Given that the ERA5 wind speed does not agree well with the in situ obervations (Sect.~\ref{s:validn}), we use the in situ MASS/DIMM observations rather than ERA5 as in the previous section. 

\begin{figure}
    \centering
    \includegraphics[width=\textwidth,trim=0cm 0cm 0cm 0cm,clip]{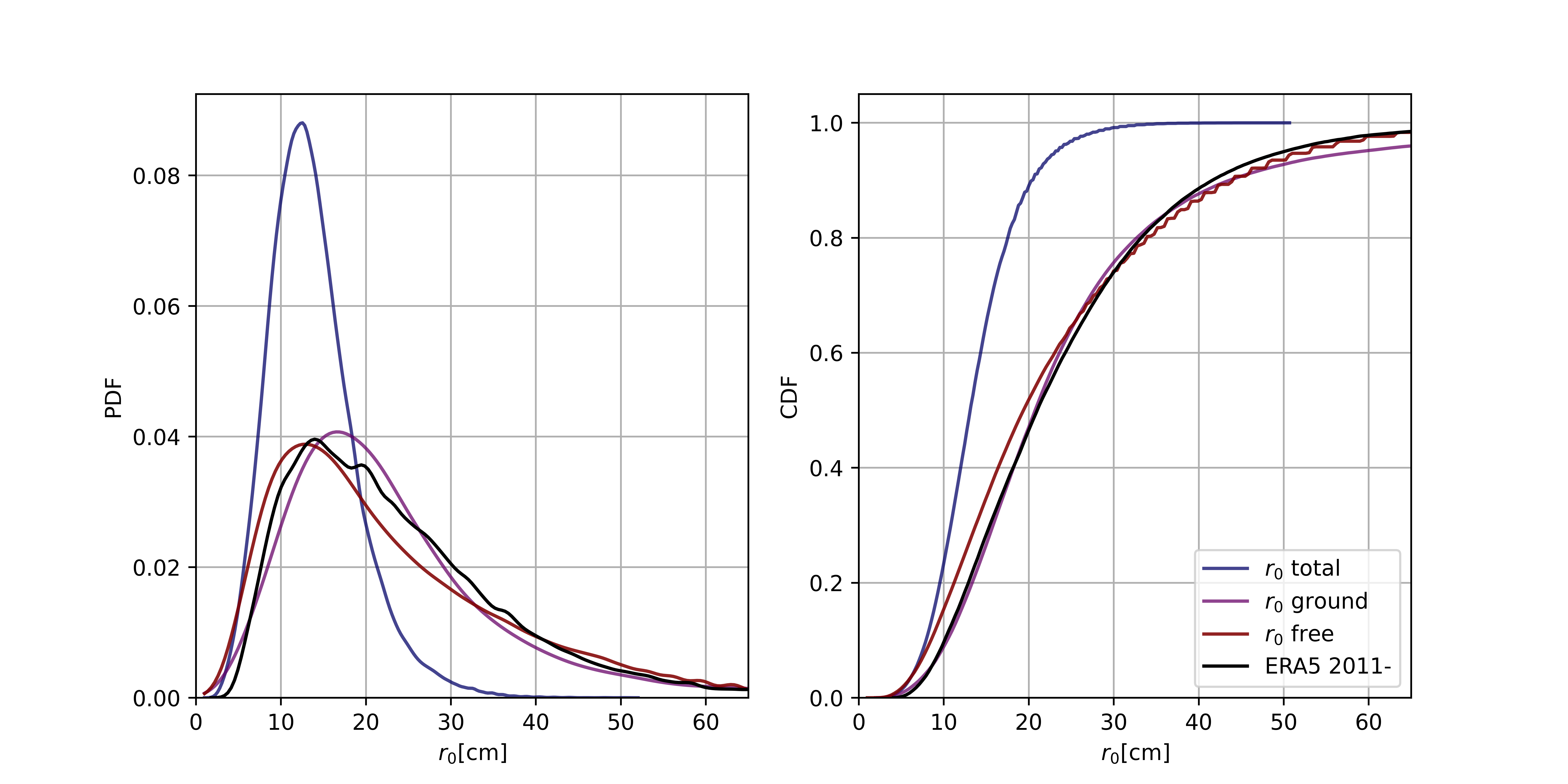}
    \caption{The total $r_0$, free atmosphere $r_0$, and ground layer $r_0$ from the MASS/DIMM and the ERA5 $r_0$ calculated from overlapping years with the MASS/DIMM. The probability density function (PDF; left) and cumulative density function (CDF; right) are shown.}
        \label{fig:fried_all}
\end{figure}

\begin{figure}
    \centering
    \includegraphics[width=\textwidth,trim=0cm 0cm 0cm 0cm,clip]{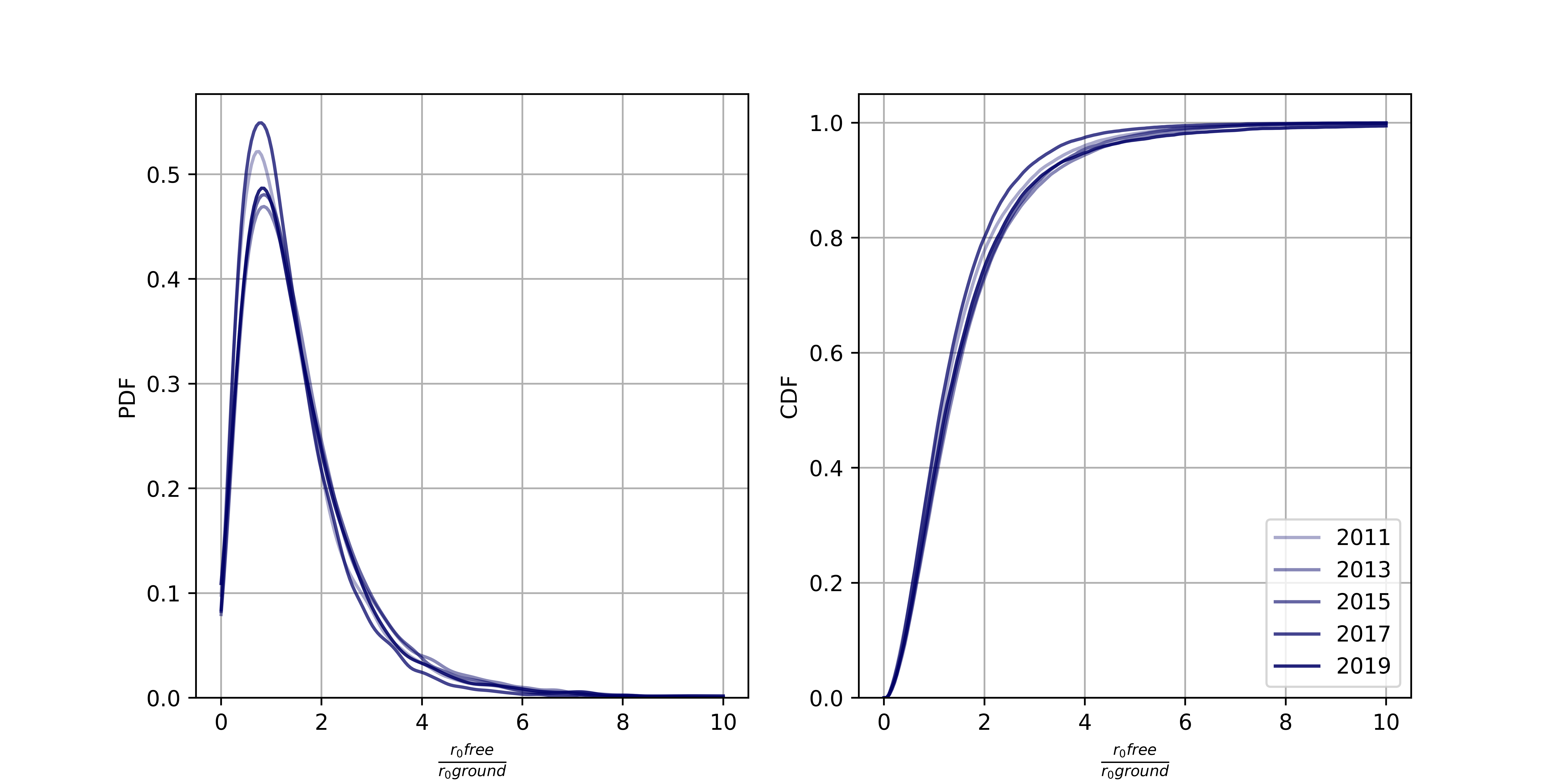}
    \caption{The probability density function (PDF; left) and cumulative density function (CDF; right) for the ratio of $r_0$ in the free atmosphere and the ground layer $r_0$ plotted for 2 year bins with the year in the legend indicating the start year.}
        \label{fig:ratio_fried}
\end{figure}

 From the MASS/DIMM measurements we not only get the full $r_0$ value covering the entire atmosphere but also the $r_0$ value of the free atmosphere. From these values we can calculate the ground layer $r_0$~\cite{Lyman_2020}. We compare the histograms of these different $r_0$ values for the complete data set in Fig.~\ref{fig:fried_all}. We see that the amount of turbulence in the ground and the free atmosphere are both log normal distributions with slightly different mean values, as expected. We also see that the bulk of the turbulence (corresponding to smaller $r_0$ values) is found in the free atmosphere. The median value of the ground layer is 21~cm which is in agreement with 20~cm what was previously found through a dedicated SLODAR campaign by Chung et al. (2009)~\cite{Chun_2009}. We also see from the figure that the ERA5 free $r_0$ and free $r_0$ have good agreement with median $r_0$ values of 19 and 22~cm, respectively. These values agree with other studies that report a 21~cm $r_0$ using MASS data~\cite{KAON303}. The mean total $r_0$ of roughly 15~cm is also in agreement with the 4-year mean of 15~cm found by Subaru~\cite{KAON303}. From Fig~\ref{fig:fried_all}, we can further verify that our calculations of the $C_n^2$ profile and $r_0$ are in good agreement with the literature. Taking a closer look we plot PDFs and CDFs for the fraction of turbulence in the free atmosphere and the ground in Fig.~\ref{fig:ratio_fried} for every 2 years. Over this short-term basis there is little evidence of a trend to either large or smaller values with the peak of the PDF fluctuating. This suggests that the strength of the ground layer does vary with respect to the free atmosphere but the mean shows no trend in time from the current baseline available. 
 
\section{\label{s:disc}Discussion}
In this section, we discuss some of the results above and their significance in relation to previous work and their implications for astronomy at the summit of Maunakea. We have chosen to highlight three key elements of our investigations, namely, 1) the trends in dome-closure criteria, 2) the impact of precipitable water vapor, and 3) turbulence, while also looking toward future data needs.

\subsection{Dome Closure Criteria}
The general increase in summit winds (Fig.~\ref{f:windchange}), although modest, is sufficient to increase the number of times Keck's dome closure criteria would be exceeded in a given year (Figs.~\ref{f:threshobs} and \ref{f:threshnights}). The overall increase in excedances is significant, with a roughly doubling of unique nights (approximately 15\% to over 30\% of nights) reporting meteorological values that could lead to dome closure over the course of a year (0.65\% per year), and a corresponding significant trend in autumn of roughly 1\% per year.

As mentioned above, it is important to note that this is not the actual dome closure rate. The thresholds are guidelines that are used by the experienced observers and telescope operators on site who are responsible for making dome open/close decisions. It is also possible that many of the ``bad nights'' would already be lost to maintenance or other non-meteorological closures that are upwards of 40 nights per year. Recent closure records\footnote{Records provided by Jim Lyke for the period spanning July 2018--October 2021} indicate that the dome is closed---for any reason---roughly 40\% of the year on average, with the lowest closure rate in May and June. Averaged historically, weather-related closures account for roughly 15\% of all closures\footnote{Q\&A at workshop in 2021 given by Keck personnel:\url{https://www.keckobservatory.org/wp-content/uploads/2021/02/OMeara-QA.pdf?x32463}}. Our analysis in Fig.~\ref{f:threshobs} agrees with these numbers provided by Keck, with lowest closure estimates in the spring and summer periods, and accounting for up to 20\% of total nighttime hours. Ultimately, this agreement lends confidence to our analysis. 

Overall, the trend is concerning. Should conditions continue to worsen, it is possible that weather-related closure could become a significant hindrance to future astronomy. We do not, however, have a long enough time series to conclude whether this is a trend that has persisted for some time, or simply a short-term increase as part of a larger cycle, and as such the conditions will improve in coming decades. We also could not take into consideration other phenomena---such as changes in cloud cover or precipitation---that would also restrict observations. It could very well be that improvements in these variables offset the worsening conditions in the variables we could consider here. Continued monitoring of the site is therefore essential. This includes the need for wider---or at least more accessible---recording and reporting on seasonal and annual dome closure statistics and their causes.

\subsection{Precipitable Water Vapor}
Taking a closer look at Fig.~\ref{f:seaspwv}, we look at the variability in the ERA5 \textit{PWV} values, specifically comparing the minima and maxima. As mentioned in Sec~\ref{s:results} following analysis of Fig.~\ref{f:pwvenso} minima in the \textit{PWV} signal follow peaks in El Ni\~no and transitions to La Ni\~no conditions. Comparing the minima of Fig.~\ref{f:seaspwv} in the years 1998, 2003, and then 2010 we see a large difference in the \textit{PWV} value and how long the dry period spans. In 2010 we have significantly longer period (twice as long) of \textit{PWV} values falling within the smallest JCMT bin. Contrasting to more recently, the \textit{PWV} values has been abnormally high providing poor conditions from 2019 until at least 2021. For astronomy such as NIR observations of exoplanets, the \textit{PWV} can significantly impact the quality of the observation and ultimately be the difference between a detection or non-detection for a given night of observation. Specifically, the discovery of a fourth planet previously undetected around HR 8799 was made around 2009 and 2010 using W.M. Keck Observatory~\cite{Marois_2010} on Maunakea when the \textit{PWV} was abnormally low for a longer period; it is possible that these conditions favorably contributed to the detection. It would be interesting to compare the time of observations for impactful science in NIR bandpass with values such as \textit{PWV} to determine how much the results depend on specific conditions in order to better understand the performance of our current and future instruments. This analysis, however, is outside the scope of this paper.

\subsection{Turbulence}
\begin{figure}
    \centering
    \includegraphics[width=0.65\textwidth,trim=0cm 0cm 0cm 0cm,clip]{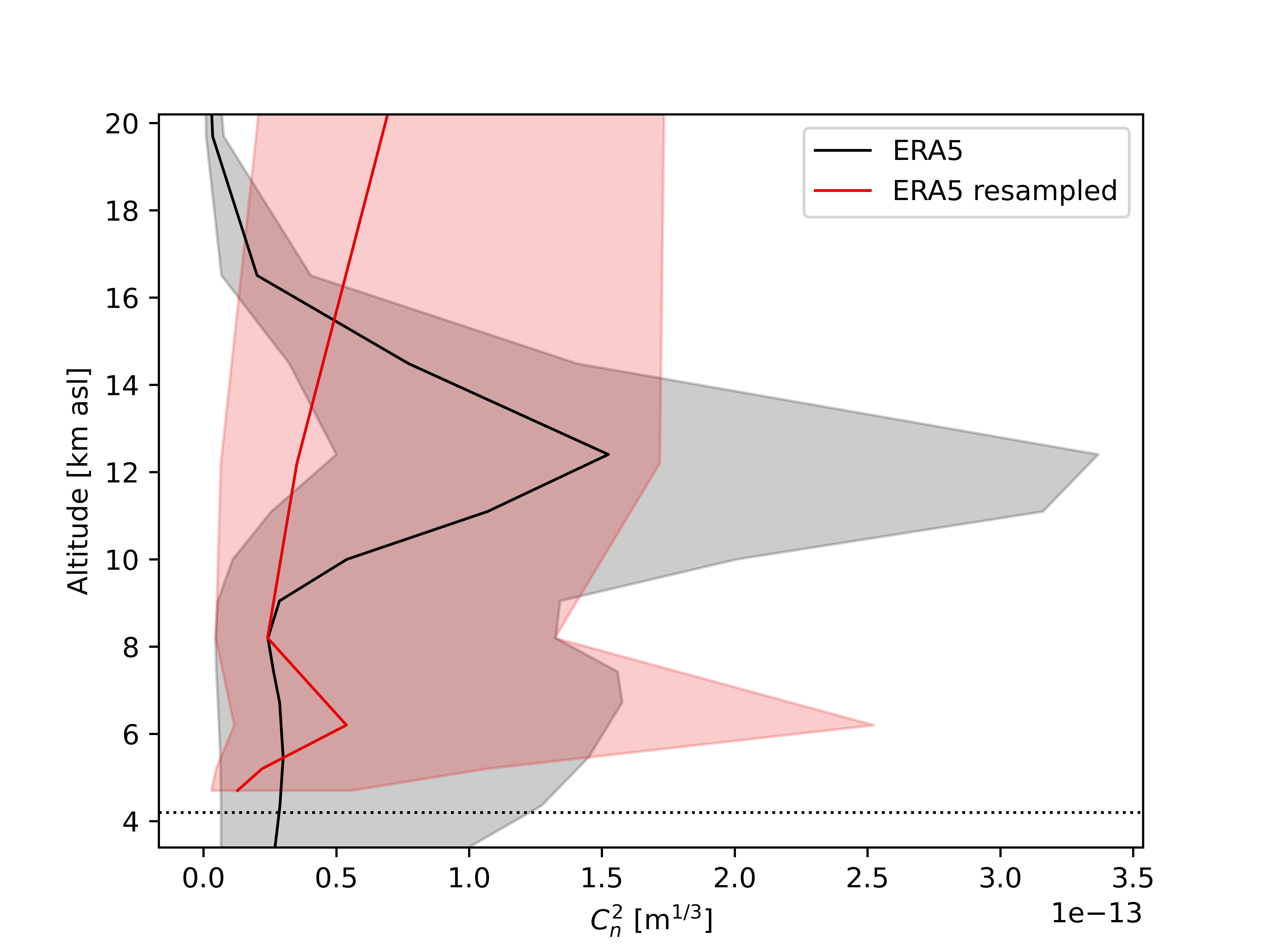}
    \caption{$C_n^2$ profile for the ERA5 at full resolution compared to ERA5 resampled to match the MASS profile. The red line is the same as in Fig.~\ref{fig:ERA5_CN2}.}
        \label{fig:era5_full}
\end{figure}

We discuss the results from Sect~\ref{s:results_turb} in more detail here. The Gladstone equation presented in Sect.~\ref{s:turbdata} depends on the shear (derivative) of the wind as a function of altitude. From the wind profiles above Maunakea (i.e., Fig.~\ref{f:meanrds}) we expect to have two peaks in the shear profile near 6~km and a second peak around 15~km (approximately where the change in wind is the greatest) on either side of the jet stream layer. When studying the CFHT $C_n^2$ profiles, however, we only find one peak around 6~km lining up with the base of the jet stream but the second peak at the top of the jet stream is not present. When resampling the ERA5 data to match the CFHT resolution we also effectively miss this second peak higher up. Figure~\ref{fig:era5_full} shows the $C_n^2$ profile for the full resolution ERA5 profile which reveals both peaks as expected. We note that the lower peak in the full profile is considerably smaller than in the resampled profile. This suggests that some of the turbulence above the jet stream is being binned into the lower layer and that the amount of turbulence is not being missed but the distribution of turbulence might be incorrect.  

MASS/DIMM instruments are model-dependent on both the assumed altitudes and the type of turbulence, providing an accuracy of up to 10\% when properly maintained~\cite{Tokovinin_2007}. Since the methods assume a thin layer of turbulence at specific heights, turbulence at different heights will be binned into a specific height. While the overall amount of turbulence measured is correct, there is much uncertainty in how it is distributed. The data suggest that the assumed layers on Maunakea for the MASS instrument could be changed to better sample where the bulk of the turbulence is expected. This might have implications for better understanding of turbulence and instrument performance but also for AO methods such a multi-conjugate AO where wavefront sensors are conjugate to different altitudes to measure the turbulence at a given height.  

With the ground layer $r_0$ showing no trend over 10 years, we comment on the impact of the increase in ground layer wind speed that is seen in the in-situ measurements presented in Sec~\ref{s:climate_mk}. The increase in strong ground layer winds not only result in dome closure criteria being met more often but will have an impact on how quickly the ground layer turbulence is evolving as well as dome seeing (turbulence inside the dome itself). More quickly flowing air over the dome structure itself (and other structures including geophysical) could alter turbulence for one telescope and not for another (i.e., for one telescope it might increase vibrations along the support structure by a small, but still significant, amount). Due to the multifaceted impact that ground layer wind can have we do not calculate the coherence time of the ground layer (and only look at free atmosphere coherence time in Sec~\ref{s:results_turb}). The full impact of change in ground-layer wind speed therefore must be evaluated for each telescope separately. 

\subsection{Future data}
As the astronomy community looks toward future telescopes such as the Thirty Meter Telescope (TMT) as well as continue to use current telescopes on Maunakea, it is desirable to expand the work presented in this paper and increase the baseline to detect trends early. With such work, new instruments and new operation methods (i.e., queue observing) can have the necessary tools and data to produce the best science with these ground-based telescopes. 

In this work we look for changes in meteorological data as well as various turbulent parameters. It is important to keep the current facilities up-to-date along with increasing their capabilities (e.g., improve MASS/DIMM resolution as well as number of operational nights). Specifically, in regards to the MASS, it will be important to understand why the distribution of turbulence is different compared to ERA5 as discussed above and make any necessary changes to what altitudes are chosen by the MASS. It would also be beneficial to have more data sources of similar data on the mountain as to not be biased towards a specific geographical feature, answering questions such as: are the winds measured at CFHT representative of winds at Keck Observatory or Subaru Telescope? Are these observatories really experiencing an increase in dome closure due to this? Finally, having observatories publish their current data on percentage of night with dome open/closed would be good for understanding how the weather is affecting astronomy and if there are indeed any trends in dome opening.

Beyond observations, numerical simulations are also important to consider. In particular, climate projections will be important in order to relate current observed trends to potential future scenarios, although careful consideration of the potential mismatch between numerical estimates and highly local in situ observations (e.g., Sect.~\ref{s:validn}) will need to be undertaken. Waiting until we can observe a change is too late. While such an analysis is beyond the scope of this current manuscript, it is an important next step.

\section{\label{s:concl}Conclusions}
We present a study of long-term trends on Maunakea with the primary goal of determining whether climate change is already having an impact on astronomy at the site. Specifically, we look at weather (temperature, wind speed, and relative humidity) both at the summit using in situ data as well as above the summit using radiosonde and re-analysis data (ERA5). We use in situ $C_n^2$ profile measurements to calibrate $C_n^2$ profile values extracted from ERA5 data allowing us to look at the turbulence characteristics over the last 40 years. 

From the meterological data we find: 
\begin{enumerate}
    \item the wind speed is increasing at the summit (Fig.~\ref{f:windchange}) with 5-year averaged speeds increasing over 30 years, 
    \item there has been a doubling in nights impacted by bad weather over the last 30 years based on the Keck dome-closure criteria (driven mainly by the wind speed), and 
    \item there is no long-term trend in precipitable water vapour \textit{PWV} although there is significant interannual variability in \textit{PWV}, possibly related to ENSO dynamics.
    \end{enumerate}
 Studying the turbulence parameters we show:  
    \begin{enumerate}
    \item that the 5-year means of $r_0$ and $\tau_0$ have not changed over the last 40 years,
    \item year to year both $r_0$ and $\tau_0$ can change noticeably, 
    \item and that the fraction of turbulence in the ground and free atmosphere has no trend in the last 10 years but note that it can vary greatly year to year. 
\end{enumerate}

To support further monitoring of climate-change impacts and to further understand the changes we are already seeing, we stress the need to maintain an up-to-data climatology on Maunakea. We would also encourage observatories to publish available data such as local temperature, wind speed, or dome closure data, making it accessible to continue this work. An important follow-up to this work is to look toward climate projections in order to better understand how climate change could affect the site in the future, and not just how it has affected Maunakea in the past (including whether any acceleration is possible). Finally, while it does not yet appear that significant deleterious changes have occurred on Maunakea, we urge the astronomy community to consider ways to reduce our carbon footprint, which will help to maintain the scientific quality of our global astronomical sites as well as the important ecological and social settings of our observatories.

\bibliography{references}
\authorcontributions{Both authors contributed equally the preparation of this manuscript.}

\funding{This research received no external funding.}

\acknowledgments{We acknowledge that the land on which the University of California, Santa Cruz is located is the unceded territory of the Awaswas-speaking Uypi Tribe. The Amah Mutsun Tribal Band, comprised of the descendants of indigenous people taken to missions Santa Cruz and San Juan Bautista during Spanish colonization of the Central Coast, is today working hard to restore traditional stewardship practices on these lands and heal from historical trauma.

The authors also wish to recognize and acknowledge the very significant cultural role and reverence that the summit of Maunakea has always had within the indigenous Hawaiian community.  We are most fortunate to have the opportunity to conduct observations from this mountain.

We are grateful to all---known and unknown---who collected and provided the data used in these analyses. Unless otherwise noted, the individual datasets are all openly available and can be accessed as described in Sect.~\ref{s:metdata} and Sect.~\ref{s:turbdata}.

This work was started while the authors were at the Leiden Observatory (MvK) and Delft University of Technology (JGI) in the Netherlands and has continued throughout the COVID-19 pandemic. We are grateful for the support received there as well as in our current positions at UC Santa Cruz. In particular, we thank Dr. Rebecca Jensen-Clem for her feedback on an early version of the manuscript. We are also grateful for the reviewer's time and feedback.
}

\end{document}